\documentclass[menuscript]{rmaa}

\newcommand{\mdeg}[2]{$#1\mbox{$^\circ \mskip-7.6mu.\,$}#2$}
\newcommand{\msec}[2]{$#1\mbox{$'' \mskip-7.6mu.\,$}#2$}

\newcommand{\Msun}{$M_\odot$}

\title{A reassessment of the kinematics of PV Cephei based on accurate proper motion measurements}

\author{Laurent Loinard, Luis F. Rodr\'\i guez
\affil{Centro de Radioastronom\'{\i}a y Astrof\'{\i}sica, UNAM, 
Morelia, M\'exico} 
        Laura G\'omez\footnote{Member of the International Max Planck
Research School (IMPRS) for Astronomy and Astrophysics at
the Universities of Bonn and Cologne.}
\affil{Max-Planck-Institut f\"ur Radioastronomie, Bonn, Germany}
        Jorge Cant\'o
\affil{Instituto de Astronom\'{\i}a, UNAM, M\'exico, DF, M\'exico}
        Alejandro C.\ Raga
\affil{Instituto de Ciencias Nucleares, UNAM, M\'exico, DF, M\'exico}
        Alyssa A.\ Goodman
\affil{Harvard-Smithsonian Center for Astrophysics, Cambridge, USA}
\and
        Hector G.\ Arce
\affil{Yale University, New Haven, USA}}

\fulladdresses{
\item Laurent Loinard and Luis F. Rodr\'\i guez: Centro de 
Radiostronom\'{\i}a y Astrof\'\i sica, Universidad Nacional 
Aut\'onoma de M\'exico, Apartado Postal 3-72, (Xangari), 58089 Morelia, 
Michoac\'an, M\'exico (l.loinard; l.rodriguez@astrosmo.unam.mx)

\item Laura G\'omez: Max-Planck-Institut f\"ur Radioastronomie, 
Auf dem H\"ugel 69, 53121 Bonn, Germany (lgomez@mpifr.de)

\item Jorge Cant\'o: Instituto de Astronom\'{\i}a, Universidad Nacional 
Aut\'onoma de M\'exico, 04510 M\'exico, DF, Mexico

\item Alejandro C.\ Raga: Instituto de Ciencias Nucleares,
Universidad Nacional Aut\'onoma de M\'exico, 04510 M\'exico, DF, Mexico
(raga@nucleares.unam.mx)

\item Alyssa A.\ Goodman: Harvard-Smithsonian Center for Astrophysics, 
60 Garden Street, Cambridge, MA 02138, USA (agoodman@cfa.harvard.edu)

\item Hector G.\ Arce: Department of Astronomy, Yale University, P.O.\ 
Box 208101, New Haven, CT 06520, USA (hector.arce@yale.edu)

} 

\shortauthor{Loinard et al.}
\shorttitle{Proper Motions of PV Cep}

\SetVolume{40} \SetFirstPage{001} \SetYear{2010}
\ReceivedDate{\today} 
\AcceptedDate{Year Month Day} 

\resumen{Presentamos dos observaciones de la estrella de pre-secuencia
principal PV Cephei tomadas con el Very Large Array con una
separaci\'on de 10.5 a\~nos. Estos datos implican que los 
movimientos propios de esta estrella son $\mu_\alpha \cos \delta$= 
+10.9 $\pm$ 3.0 mas a\~no$^{-1}$; $\mu_\delta$ = +0.2 $\pm$ 1.8 mas 
a\~no$^{-1}$, muy similares a los ya conocidos para HD 200775, la 
estrella B2Ve que domina la iluminaci\'on de la cercana nebulosa de 
reflexi\'on NGC~7023. Este resultado sugiere que PV Cephei no es una estrella desbocada con movimientos r\'apidos,
como se hab\'\i a sugerido en anteriores estudios. La alta velocidad de
PV Cep se hab\'{\i}a inferido de los desplazamiento sistem\'aticos hacia 
el este de los bisectores de pares sucesivos de objetos Herbig Haro a lo largo 
de su chorro. Estos desplazamientos sistem\'aticos podr\'{\i}an, en realidad,
deberse a una disimetr\'{\i}a en los mecanismos de eyecci\'on, o a 
una asimetr\'{\i}a en la distribuci\'on del material circunestelar.}
 
\abstract{We present two Very Large Array observations of the
pre-main-sequence star PV Cephei, taken with a separation of 10.5
years. These data show that the proper motions of this star are 
$\mu_\alpha \cos \delta = +10.9 \pm 3.0$ mas yr$^{-1}$; 
$\mu_\delta = +0.2 \pm 1.8$ mas yr$^{-1}$, very similar to those --
previously known-- of HD 200775, the B2Ve star that dominates 
the illumination of the nearby reflection nebula NGC~7023. This 
result suggests that PV Cephei is not a rapidly
moving run-away star as suggested by previous studies. The large
velocity of PV Cephei had been inferred from the systematic eastward
displacement of the bisectors of successive pairs of Herbig Haro knots along 
its flow. These systematic shifts might instead result from an intrinsic 
dissymmetry in the ejection mechanisms, or from an asymmetric
distribution of the circumstellar material.}


\begin{document}

\maketitle

\section{Introduction}

Compact stellar clusters as well as multiple stellar systems with
three or more members can be dynamically unstable (e.g.\ Valtonen \&
Mikkola 1991, Poveda et al.\ 1967). Indeed, close few body encounters occurring within them
can lead to the acceleration of one or more of the system members,
which --if the acceleration is sufficient-- can escape their
birthplace and become run-away stars. It is important to search for
direct observational evidence of these energetic events, because if
they occurred with sufficient frequency during the earliest stages of
star-formation, they could have a significant impact on the very
outcome of the star-forming process (Reipurth 2000). Arguably the most
promising case so far, is that of the BN/KL region in Orion where
three stars (including BN itself) are moving away at about 30 km
s$^{-1}$ from a common point of origin, which happens to be near the
center of the massive KL outflow (Rodr\'{\i}guez et al.\ 2005, G\'omez
et al.\ 2005, 2008, Zapata et al.\ 2009). The dynamical disruption in that case, 
appears to have occurred a mere 500 years ago.

Goodman \& Arce (2004 --hereafter GA2004) have suggested that the
star PV Cephei (hereafter PV Cep) might be another potential
candidate. PV Cep (Tab.\ 1) is a Herbig Ae/Be star (Li et al.\ 1994) that drives
a well-studied giant Herbig Haro (HH) flow (Reipurth et al.\ 1997; G\'omez et al.\
1997; Arce \& Goodman 2002a, 2002b; GA2004). It is located about
\mdeg{1}{5} west of the reflection nebula NGC~7023, and the centroid
velocity of its CO emission is similar to that of the CO emission from
NGC~7023 itself (Cohen et al.\ 1981). Because of this shared
kinematics, it is very likely that PV Cep and NGC~7023 are at the same
distance from the Sun. The measured trigonometric parallax of
HD~200775 (MWC 361, V = 7.4 mag) the B2Ve star that dominates the
illumination of NGC~7023 (Herbig 1960) yields a distance for the
entire region of 430$^{+155}_{-91}$ pc (Perryman et al.\ 1997). This
value is consistent with the somewhat older estimate of 500 pc used by
GA2004. NGC~7023 is associated with a compact cluster of young stars,
that contains a number of low luminosity pre-main-sequence objects
(Lepine \& Rieu 1974).

\begin{table*}[!t]
\small
\caption{Characteristics of PV Cep}
\begin{center}
\begin{tabular}{ll}
\hline
\hline
Spectral type & A5 \\%
R-band magnitude & $R$ = 11.1 \\%
Distance & $d$ = 430$^{+155}_{-91}$ pc \\%
Mean position & $\alpha$ = $20^h 45^m 53\rlap.^{s}9550 \pm 0\rlap.^{s}0014$  \\%
(Equinox J2000, Epoch 2002.26) & $\delta$ = $67^\circ 57{'} 38\rlap.{''}681 \pm 
0\rlap.{''}008$ \\
\hline
\hline
\label{tab:2}
\end{tabular}
\end{center}
\end{table*}

The suggestion by GA2004 that PV Cep might be a run-away star 
was based on a careful analysis of the kinematics and morphology 
of the gas surrounding PV Cep. The strongest argument follows from 
a study of the distribution of successive pairs of HH knots located at 
nearly equal distances from PV Cep along its jet. The segments 
joining such successive pairs of knots are expected to pass through 
the source driving the flow --if the latter is stationary. In the case of 
PV Cep, however, GA2004 found that the bisectors of the segments 
were systematically shifted eastward of the source, and increasingly 
more so, as older HH pairs were considered. This suggests that the 
source used to be somewhat to the east of its current position, and 
thus, that it is moving westward relative to its surroundings. Assuming
that the HH knots along the flow are fully detached clumps, GA2004 
estimate the velocity of the central source to be 22 km s$^{-1}$ (see 
discussion in Sect.\ 2). Most remarkably, the velocity vector obtained 
by GA2004 points almost exactly {\em away} from NGC~7023, as if PV 
Cep had been ejected from NGC~7023 about 10$^5$ yr ago (see their 
Fig.\ 6). Thus, PV Cep might indeed provide us with an example of a 
run-away star dynamically ejected from its (still identifiable) parent 
cluster. Note, however, that the conclusions of GA2004 are based on 
an analysis of the kinematics and morphology of the gas surrounding 
PV Cep, and {\em not} on a direct measurement of the relative proper 
motion between PV Cep and NGC~7023. In the present paper (Sect.\ 
3 and 4), we will provide such a direct measurement combining 
published optical proper motions of the brightest star in NGC~7023 
and new radio measurements of the proper motion of PV Cep itself. 

\begin{figure*}[!t]
\centering
\includegraphics[scale=0.55]{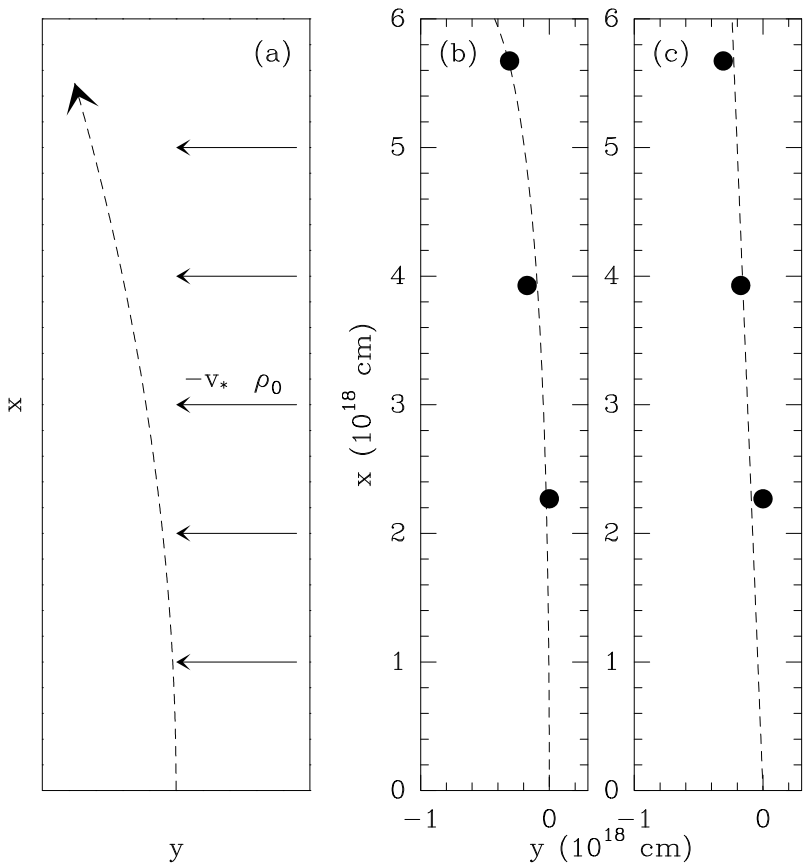}
\caption{(a) Schematic diagram of a star which ejects a bipolar outflow 
along the positive $x$-axis, in a direction perpendicular to the motion 
of the star with respect to the surrounding environment. In the frame of 
reference of the star, the environment moves towards 
the left along the $y$-axis, producing a curvature in the bipolar outflow 
system. (b) and (c) The black filled circles show the positions of the 
ejecta relative to the source according to the results of GA2004, and
the dashed lines represent the best fit for fully detached clumps (panel
(b); see Sect.\ 2.1) and for a continuous jet (panel (c); see Sect.\ 2.2).}
\label{fig:models}
\end{figure*}   

\section{A re-analysis of the bisector results}

Several theoretical models (both analytic and numerical) have been
proposed to explain the curved morphology presented by some HH jets. 
Cant\'o \& Raga (1995) have developed an analytical model that treats 
the problem of a {\em steady} jet interacting with a sidewind. More 
recently, Masciadri \& Raga (2001) presented numerical simulations 
of a {\em variable} jet in a sidewind and identified two different regimes. 
For low-amplitude variability (caused by moderate changes in the jet 
ejection velocity), small working surfaces (i.e.\ HH knots) are created 
along the jet, and they travel following the path predicted by the steady 
jet model of Cant\'o \& Raga (1995). On the other hand, large working 
surfaces created by larger changes in the jet ejection velocity eventually 
become detached ``clumps'' that travel on somewhat straighter
trajectories. 

\subsection{Analytic solution for fully detached clumps}

In their analysis of the bisectors of successive HH pairs, GA2004 assumed 
that the ejecta had become detached clumps, and integrated the equations 
of motion numerically to obtain the trajectory of the clumps. Recently, however, 
Cant\'o et al.\ (2008) showed that the idealized case of fully detached clumps 
has a completely analytic solution, which produces trajectories similar to 
those obtained in numerical simulation. To examine that analytic solution in 
the case of PV Cep, we assume (following GA2004) that PV Cep is moving 
relative to the ambient gas in a direction exactly perpendicular to that of its jet. 
The velocity of the star relative to the surroundings is taken to be $v_*$, and 
individual clumps of mass $M$ are ejected at a velocity $v_e$ (see Fig.\ 
\ref{fig:models}a for a schematic description). The trajectory 
is parametrized by a distance $s_0$ (Cant\'o et al.\ 2008) given by:

\begin{equation}
s_0 \equiv {3 \over 4} \left( {M v_0^4 \over \xi \rho_0 c^4} \right)^{1/3}.
\end{equation}

\noindent Here, $\xi$ is a constant of integration whose numerical 
value is $\xi$ = 14 (Cant\'o et al.\ 1998), $c$ is the isothermal sound
speed of the clump material, and $\rho_0$ is the density of the ambient 
gas. In the present case where $v_*$ and $v_e$ are perpendicular,
$v_0$ is simply 

\begin{equation}
v_0 = \sqrt{v_*^2+v_e^2}. 
\end{equation}

\noindent If one further defines the angle $\theta$ as

\begin{equation}
\tan \theta = {v_e \over v_*},
\end{equation}

\noindent the equation of the trajectory is given analytically by:

\begin{equation}
x = {y \over \tan \theta}-4s_0\left( {v_* \over v_0} \right) \left[ 1- \left( 1-{y \over s_0 \sin \theta}\right)^{1/4}\right].
\end{equation}

\noindent
Following GA2004, we adopt  $v_e$ = 350 km s$^{-1}$ and $\rho_0$ = 1.5 $\times$ 
10$^3$ cm$^{-3}$. A good fit to the data points (shown in Fig.\ \ref{fig:models}a) is obtained 
for $v_*$ = 22 km s$^{-1}$, provided that $s_0$ $\approx$ 6.4 $\times$ $10^{18}$ cm.
This, in turn, implies $M$ $\approx$ 2 $\times$ 10$^{-4}$ \Msun\ if the value
for $c$ is taken to be 3.2 km s$^{-1}$ as suggested by GA2004. A comparison 
between the analytic trajectory obtained using these parameters and the path 
obtained numerically by GA2004 (see their Fig.\ 2) shows that the analytical and 
numerical trajectories agree extremely well.

It should be mentioned that $c$ in equation (1) is the isothermal sound speed {\em 
of the ejecta}. Such ejecta are believed to be ionized, so the appropriate sound 
speed is $c$ $\approx$ 10 km s$^{-1}$. In their analysis, however, GA2004 
used $c$ = 3.2 km s$^{-1}$, the typical velocity dispersion of the {\em ambient} 
molecular gas. From equation (1), it is clear that to obtain a similar value of $s_0$ for 
$c$ = 10 km s$^{-1}$ without resorting to unreasonable assumptions for $\rho_0$ 
and $v_0$, one has to increase the mass of the ejecta by nearly two orders of
magnitude, to about 1.5 $\times$ 10$^{-2}$ \Msun. Even if one allows $v_e$ 
to increase to 400 km s$^{-1}$ and $\rho_0$ to decrease to 1000 cm$^{-3}$, the 
ejecta must still have masses of order 10$^{-2}$ \Msun. Moreover, the transverse 
velocity of the star must then be increased to about 40 km s$^{-1}$ to obtain a
good fit. We conclude that if the ejecta from PV Cep are indeed fully detached 
clumps of ionized gas, they must be very massive\footnote{As a consequence,
in this scenario, PV Cep must have undergone episodes of extremely high 
accretion/ejection.}, and the velocity of PV Cep might be even larger that the 
22 km s$^{-1}$ proposed by GA2004.

\subsection{The HH knots of PV Cep as part of an underlying continuous jet}

The trajectories considered in Sect.\ 2.1 appropriately describe jets in which 
large amplitude perturbations leading to fully detached clumps have developed. 
As mentioned earlier, lower amplitude perturbations create HH knots that travel 
along different paths: those predicted by the analytical model of Cant\'o \& Raga 
(1995). Moreover, the analysis of Masciadri \& Raga (2001) shows that a smaller 
sidewind velocity is necessary to create a given jet curvature if the HH knots are 
such small perturbations along an underlying continuous jet than if they are fully 
detached clumps. Well-defined HH knots suggestive of detached clumps are 
certainly present along the jet driven by PV Cep (GA2004 and references therein). 
There is also evidence, however, for an underlying continuous jet which is 
always ``on''. Indeed, the very existence of steady, compact centimeter radio emission 
associated with PV Cep (which traces free-free emission from the base of an 
ionized flow; see Sect.\ 3 below) demonstrates that such a continuous jet exists and 
that the ejection of matter is not fully episodic. The detection of water masers 
(Marvel 2005) and of high velocity entrained molecular gas (Arce \& Goodman
2002b) in the immediate vicinity of PV Cep confirm the existence of such
a steady jet. Let us, therefore, estimate the sidewind velocity that would
be required to explain the bisector results of GA2004 in a steady wind scenario.
The applicability of this model to the specific case of PV Cep will be discussed
in Sect.\ 5.

As shown by Cant\'o \& Raga (1995), in the vicinity of the source, a bipolar flow in 
a sidewind approximately follows a parabolical shape\footnote{Farther 
out, the jet move on nearly straight trajectories aligned with the direction of the side 
wind (Cant\'o \& Raga 1995). Since the knots in PV Cep are clearly not moving 
perpendicularly to the direction of the jet (GA2004), the portion of the jet considered
here can be assumed to be parabolic.}  (with both the jet and the counterjet 
on the same parabola) described by

\begin{equation}
y={x^2\over 2\lambda}\,,
\label{zy}
\end{equation}

\noindent where $x$ is parallel to the initial outflow direction, and
$y$ is the coordinate perpendicular to $x$, and parallel to the motion
of the source (Fig.\ \ref{fig:models}a). This relation is valid for the case 
in which the motion of the source is perpendicular to the ejection.
The distance $\lambda$ in Equation (\ref{zy}) is given by

\begin{eqnarray}
\lambda & \equiv & \left({{\dot M} v_e^3\over \pi c^2\rho_0 v_*^2}\right)^{1/2} \nonumber \\
& = & {\rm 7.8\times 10^{-18}\,cm}
\left({{\dot M}\over 10^{-7}{\rm ~M}_\odot{\rm yr}^{-1}}\right)^{1/2}
\left({v_e\over 100\,{\rm ~km\, s}^{-1}}\right)^{3/2}
\left({\rho_0\over 10^3{\rm ~cm}^{-3}}\right)^{-1/2}
\left({v_*\over 1\,{\rm ~km\,s}^{-1}}\right)^{-1}\,,
\label{lam}
\end{eqnarray}

\noindent where $\dot M$ is the mass-loss rate, and the other parameters have the 
same meaning as before. The best fit to the observed positions reported by GA2004 
using equation (\ref{zy}) yields $\lambda=4.64\times 10^{19}$~cm.  With this value, 
we can now use equation (\ref{lam}) to determine the required velocity for the environment:

\begin{equation} \left({v_*\over 1\,{\rm km\,s}^{-1}}\right) =0.17
\left({{\dot M}\over 10^{-7}{\rm ~M}_\odot{\rm yr}^{-1}}\right)^{1/2}
\left({v_e\over 100\,{\rm ~km\, s}^{-1}}\right)^{3/2} \left({\rho_0\over
10^3{\rm ~cm}^{-3}}\right)^{-1/2} \label{va}. \end{equation} 

\noindent
Finally, using the values $v_e=350$~km~s$^{-1}$ and $\rho_0=1500$~cm$^{-3}$
determined by GA2004, we obtain $v_*=0.9$~km~s$^{-1}$, assuming a mass 
loss rate ${\dot M}=10^{-7}$~M$_\odot$yr$^{-1}$. The corresponding trajectory
is shown in Fig.\ \ref{fig:models}c.

Thus, a very modest relative motion --of the order of the typical 
stellar velocity dispersion in newborn stellar clusters-- between PV Cep 
and its surroundings would be required to explain the bisector
results of GA2004 in the case of a continuous jet. We will return
to this point in Sect.\ 5.

\begin{figure*}[!t]
\centering
\includegraphics[scale=0.34]{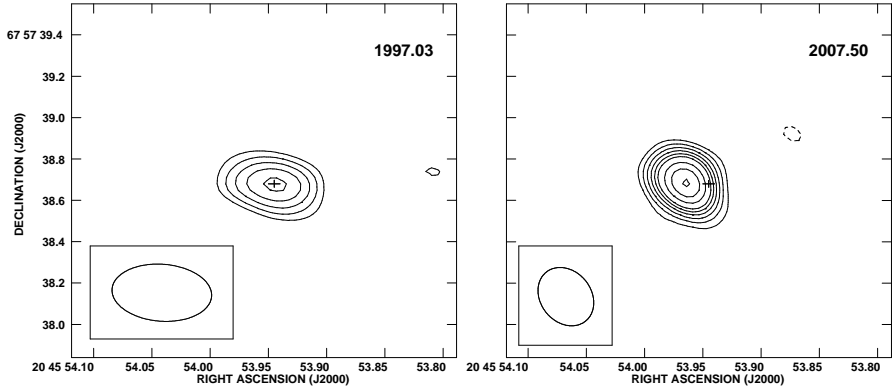}
\caption{VLA 3.6 cm images of PV Cep for 1997.03 (left) and 2007.50 
(right). The contours are -3, 3, 4, 5, 6, 7, 8, 10, 12 and 14 times
19 and 12 $\mu$Jy beam$^{-1}$, the rms noise of the first and second
epochs, respectively. The cross marks the position and positional 
error at the 1997.03 epoch. The half-power synthesized beam for each 
epoch ($0\rlap.{''}48 \times 0\rlap.{''}28$ ; PA = $+86^\circ$ for 
1997.03 and $0\rlap.{''}30 \times 0\rlap.{''}24$ ; PA = $+39^\circ$
for 2007.50) is shown in the bottom left corner of the images.}
\label{fig:data}
\end{figure*}

\section{Observations}

In this paper, we will make use of 3.6 cm continuum observations 
obtained on 1997 January 10 and 2007 July 2 with the Very Large
Array (VLA) of the NRAO\footnote{The National Radio Astronomy
  Observatory is operated by Associated Universities Inc. under
  cooperative agreement with the National Science Foundation.}  in its
most extended (A) configuration. This combination of wavelength and
configuration provides an angular resolution of about $0\rlap.{''}3$
for images with natural weighting.  The absolute amplitude calibrator
was 1331+305 (with an adopted flux density of 5.20 Jy at 8.4 GHz)
and the phase calibrator was 2022+616, with bootstrapped flux
densities of 3.36$\pm$0.03 Jy and 3.07$\pm$0.02 Jy, for the first and
second epochs, respectively. For the first epoch only 14 of the 27
antennas of the VLA were used, with an on-source integration time of
4.2 hours. For the second epoch we used 23 of the 27 antennas of the
VLA, with an on-source integration time of 3.2 hours.  The data were
edited and calibrated using the Astronomical Image Processing System
(AIPS) software package (Greisen 2003).

To ensure accurate astrometry, the position of the calibrator 2022+616
for the 1997 observations was updated to the refined position given in
the VLA catalog of calibrators for 2007. The continuum images shown in
Fig.\ \ref{fig:data} were used to measure the positions and flux densities of PV
Cep for the two epochs (Tab.\ 2). The source appears to be unresolved
at both epochs, and the emission is most probably free-free radiation
from ionized gas at the base of the outflow driven by PV Cep (i.e.\
Eisl\"offel et al. 2000). The small variation in flux density between
the two epochs is common for this type of sources (e.g.\
Galv\'an-Madrid et al.\ 2004), and most likely reflects moderate
variability either of the mass-loss rate or of the velocity of the
wind.  

As can be seen in Fig.\ \ref{fig:data} and Tab.\ 2, the source shows a small
displacement ($\sim$ \msec{0}{11}) toward the east. The proper 
motions of the source measured over the 10.47 year time baseline of the 
observations are $\mu_\alpha \cos \delta = +10.9 \pm 3.0$ mas 
yr$^{-1}$; $\mu_\delta = +0.2 \pm 1.8$ mas yr$^{-1}$. The base
of the jet driven by PV Cep (as traced by the radio emission) is 
expected to follow closely the motion of the star itself, because
the emission is largely dominated by the highest density gas
in the wind, which is located very close to the star since the
gas density typically falls as $n(r) \propto r^{-2}$. The high 
accuracy with which VLA observations of free-free emission 
from young stars can trace the stellar motions themselves can 
be seen, for instance, from the multi-epoch observations of 
IRAS~16293--2422 (Loinard et al.\ 2007, Pech et al.\ 2010) or 
of T Tau N (Loinard et al.\ 2003). Thus, the proper motions 
measured here can safely be assumed to represent the motion of 
PV Cep. 

\begin{table*}[!t]
\small
\caption{Positions and Flux Densities for PV Cep}
\begin{center}
\begin{tabular}{lcccc}
\hline
\hline
&\multicolumn{2}{c}{Position} & 3.6-cm Flux \\
\cline{2-3} 
Epoch &  $\alpha$(J2000) & $\delta$(J2000) & Density (mJy) \\
\hline
1997.03 & $20^h 45^m 53\rlap.^{s}9449 \pm 0\rlap.^{s}0051$ & 
$67^\circ 57{'} 38\rlap.{''}680 \pm 0\rlap.{''}017$ & $0.14\pm0.02$ \\
2007.50 & $20^h 45^m 53\rlap.^{s}9651 \pm 0\rlap.^{s}0014$ & 
$67^\circ 57{'} 38\rlap.{''}682 \pm 0\rlap.{''}008$ & $0.18\pm0.01$ \\
\hline
\hline
\label{tab:1}
\end{tabular}
\end{center}
\end{table*}

\section{Results}

To test the proposal of GA2004 that PV Cep might be moving at $\sim$ 22
km s$^{-1}$ away from NGC~7023, one must compare the proper motion of
PV Cep derived above with the proper motion of stars within
NGC~7023. The only star of NGC~7023 (and indeed, the only star within
a few degrees of PV Cep) for which we found reliable published proper
motions is HD~200775, the bright star responsible for most of the
illumination of NGC~7023 (see Sect.\ 1). Since it is the brightest member
of the cluster, HD~200775 is likely to be very near its center of
mass, and its proper motion should provide a good estimate of the
proper motion of the entire cluster. The most recent reported proper
motion for HD~200775 can be found in Ducourant et
al. (2005)\footnote{Indeed, Ducourant et al. (2005) also report proper
  motions for PV Cep. However, their values of $\mu_\alpha \cos
  \delta$ = $-$3 $\pm$ 7 mas yr$^{-1}$; $\mu_\delta$ = $-$14 $\pm$ 7
  mas yr$^{-1}$ have significantly larger errors than our
  measurements.}; they give $\mu_\alpha \cos \delta = +11 \pm 2$ mas
yr$^{-1}$; $\mu_\delta = +2 \pm 2$ mas yr$^{-1}$. HD~200775 is also
contained in the Hipparcos catalog, where the reported proper motions
are $\mu_\alpha \cos \delta = +6.74 \pm 0.64$ mas yr$^{-1}$;
$\mu_\delta = -1.48 \pm 0.54$ mas yr$^{-1}$. These two values are
consistent with one another at the 1.5$\sigma$ level.

The relative motion between PV Cep and NGC~7023 (obtained by subtracting
the proper motion of PV Cep measured above from that of NGC~7023) is

\[ \Delta ( \mu_\alpha \cos \delta) = +0.1 \pm 3.6 \mbox{~mas yr}^{-1}  ~~~~~~~~~~~  \Delta ( \mu_\delta) = +1.8 \pm 2.7 \mbox{~mas yr}^{-1}, \]

\noindent if the values quoted by Ducourant et al.\ (2005) are used. 
Using the Hipparcos figures, we obtain:

\[ \Delta ( \mu_\alpha \cos \delta) = - 4.2 \pm 3.1 \mbox{~mas yr}^{-1}  ~~~~~~~~~~~  \Delta ( \mu_\delta) = -1.7 \pm 1.9 \mbox{~mas yr}^{-1}. \]

\noindent Recall that GA2004 predicted a relative velocity of 22 km
s$^{-1}$ toward the west, corresponding to a relative proper motion
(almost entirely in right ascension) of $\Delta ( \mu_\alpha \cos
\delta) = + 9.3 \mbox{~mas yr}^{-1}$. The observed values are
inconsistent at the 3$\sigma$ level with this prediction,\footnote{It is, of 
course, even more inconsistent with the relative velocity of 40 km s$^{-1}$
obtained in the detached clump scenario assuming a sound speed of
10 km s$^{-1}$ (see Sect.\ 2.1).} and show that PV Cep is in fact very nearly stationary relative to NGC~7023 
both in right ascension and declination. If anything, the Hipparcos 
value suggests a marginal motion of PV Cep {\it toward} NGC~7023. 
We conclude that PV Cep is very unlikely to have been ejected from 
NGC~7023 about 10$^5$ yr ago.

Strictly speaking, the analysis of GA2004 does not require PV
Cep to be moving away from NGC~7023 at 22 km s$^{-1}$. It only
requires PV Cep to be moving at that speed relative to its surrounding
gas. The morphology of the jet driven by PV Cep could still be
explained by the model proposed by GA2004 if the gas cloud in which PV
Cep is embedded were moving at 22 km s$^{-1}$ {\em toward the east}
relative to both PV Cep and NGC~7023. We consider this possibility
very unlikely, however, because (i) as mentioned in Sect.\ 1, the radial
velocity of the gas associated with NGC~7023 and with PV Cep are very
similar, and (ii) it is not clear what physical mechanism could
accelerate an entire molecular cloud to 22 km s$^{-1}$, particularly
without also accelerating the stars that formed within it.

\section{Discussion}

Given that a relative velocity of 22 km s$^{-1}$ between PV Cep and
its surroundings appears to be inconsistent with the present observations,
one must seek a different interpretation of the systematic eastward 
shift of the bisectors of successive HH pairs along the flow driven by 
PV Cep. One obvious alternative possibility is provided by the continuous jet 
model presented in Sect.\ 2.2. In that scenario, a relative velocity between 
PV Cep and NGC~7023 of a few km s$^{-1}$ would be sufficient to explain 
the bisector data, and such a small relative velocity would be fully consistent 
with the astrometry presented in Sect.\ 3. We note, indeed, that continuous 
jet models have been used to explain the curved morphologies of  other 
young stellar flows. Of particular relevance is the case of the jet
driven by HH 30 (L\'opez et al.\ 1995) whose curved shape was interpreted 
by Cant\'o \& Raga (1995) in terms of such a continuous flow in a sidewind.

It is important to note, however, that the flow driven by PV Cep does show 
clear signs of strong variability (GA2004 and references therein). A steady 
jet is clearly present in the system, but the knots considered by GA2004 in 
their bisector analysis have likely been created during episodes of significantly 
increased mass-loss.\footnote{Note that the flow in HH 30 appears overall 
more continuous than that in PV Cep, and does lend itself more naturally to 
a treatment based on the hypothesis of a steady jet.} To completely describe
that situation, one should consider a steady, low-level (i.e.\ moderate mass-loss 
rate) underlying jet, periodically undergoing episodes of enhanced mass-loss. 
This is a case intermediate between the two idealized descriptions provided in 
Sect.\ 2, so numerical simulations would in principle be needed to treat it in 
general. However, since the periodic mass-loss enhancements undergone by 
PV Cep seem particularly strong, the correct description for that specific source 
is almost certainly more similar to the case of detached bullets than to that of a 
continuous jet. In this situation, a relative velocity between PV Cep and its
surroundings of at least 20 km s$^{-1}$ would be needed to explain the
bisector results of GA2004, and that appears to be inconsistent with our new
measurements. 

\medskip

Another alternative interpretation of the bisector results of GA2004 would be to 
invoke asymmetries. The basic assumption behind the argumentation of GA2004 
is that pairs of clumps ejected from a stationary star propagate symmetrically away 
from it, in such a way that the bisectors of such ejecta pairs intercept the driving 
source. This assumption would be violated, however, if the ejecta 
on the two sides of the star moved away at different speeds and/or not
exactly in opposite directions, either because of an intrinsic dissymmetry 
in the jet launching mechanism or because of an asymmetry in the density of the 
circumstellar material. Several jets (e.g.\ HH 1--2. Cepheus A, or HH 80; 
Rodr\'{\i}guez et al.\ 2000, Mart\'{i} et al.\ 1998, Curiel et al.\ 2006) are known to 
exhibit this kind of asymmetries. Arguably the clearest case is that of the very 
young low-mass protostar IRAS~16293--2422 (Pech et al.\ 2010). 
In this object, a bipolar pair of clumps has been ejected just a few years ago, 
and each ejecta has now moved (in projection) about 60 AU away from the central 
source. Multi-epoch radio observations have shown that the projected 
speed of the southern ejecta is two to three times higher than that of the 
northern clump and that --in at least one of the observations-- the line joining the 
ejecta does not pass through the position of the driving source (Pech et al.\
2010, particularly their Fig.\ 2). In addition, the large-scale molecular outflow powered 
by that jet system is known to be highly asymmetric (e.g.\ Castets et al.\ 2001, Hirano 
et al.\ 2001), the northern lobe being much more conspicuous, massive, and 
extended that its southern equivalent. Interestingly, the molecular outflow in PV Cep 
happens to be equally asymmetric (e.g.\ Arce \& Goodman 2002a) with a northern lobe 
about 2.5 times more massive and 1.5 times more extended than its southern counterpart.
Could an asymmetry similar to that in IRAS~16293--2422 be at the origin of the PV Cep 
bisector results of GA2004?

Three pairs of HH groups have been identified along the jet driven by PV Cep: 
HH~315 A, B, and C towards the north-west, and HH~315 D, E, and F toward 
the south-east (e.g.\ G\'omez et al.\ 1997, see Fig.\ 3). They are believed to 
have been created during three episodes of enhanced accretion/ejection
(G\'omez et al.\ 1997, GA2004, Arce \& Goodman 2002a). Moreover, these HH 
groups are not distributed along a straight line, but along an S-shaped
structure, clearly resulting from precession in the underlying jet (Fig.\ 3). A simple
precessing jet model that reasonably accounts for that structure has been 
proposed by G\'omez et al.\ (1997) and is reproduced in Fig.\ 3a. This model
(e.g.\ Hjellming \& Johnston 1981) assumes that the jet axis precesses on the 
surface of a cone of opening angle $\theta$, inclined by an angle $i$ from the 
line of sight. The jet has constant ejection velocity $v_e$ and precession 
velocity $\omega$, and the position angle of the precession axis is $\phi$.
The black line on Fig.\ 3a corresponds to $\theta$ = 22.5$^\circ$, $i$ = 80$^\circ$,
$\omega$ = 1.5 arcmin yr$^{-1}$, v = 180 km s$^{-1}$, and $\phi$ = --36.5$^\circ$
(G\'omez et al.\ 1997). As discussed by G\'omez et al.\ (1997), this model shows
that successive episodes of enhanced mass-loss occurred in PV Cep roughly 
every 2,000 years. 

\begin{figure*}[!t]
\centering
\includegraphics[scale=0.5,angle=-90]{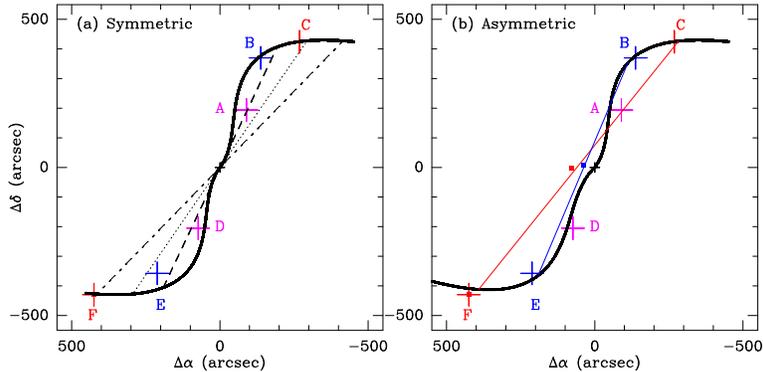}
\caption{Distribution of groups of HH knots around PV Cep. The
three groups towards the north-west are labelled A, B, and C, and
the three groups toward the south-east are D, E, and F. The positions
are taken from G\'omez et al.\ (1997). (a) Model based on a
symmetric precessing jet following G\'omez et al.\ (1997) with the parameters 
given in the text (solid black line). Lines joining the expected positions along 
the jet at times 6,000, 7,000, and 8,000 yr are shown. (b) Model based on an 
asymmetric precessing jet with the parameters given in the text (solid black line). 
Lines joining the expecting positions along the jet at 4,900 and 6,900 
yr are shown. They intercept the positions of the bisectors of the B-E
and C-F pairs (shown as blue and red squares).}
\label{fig:precess}
\end{figure*}

An asymmetry in the distribution of the HH groups is immediately
apparent in Fig.\ 3a. For instance, while the C-F pair is expected to have been
created during the same enhanced mass-loss episode, the symmetric
model shown in Fig.\ 3a would ascribe them ages differing by about 1,000 yr.
A similar situation occurs for the B-E pair. Of course, this age inconsistency is
simply a different way of re-expressing the bisector conclusions of GA2004. 
Because of the asymmetry in the distribution of the HH groups, the segments 
joining C to F and B to E do not pass through the current position of the driving 
source. The situation can be significantly improved by relaxing slightly the symmetry 
of the jet. As an example, we show in Fig.\ 3b a model identical to that of G\'omez 
et al.\ (1997) for the northern part of the flow, but slightly different for the 
southern counterpart. Specifically, the velocity of the southern jet is 10\% larger 
than that of the northern jet, and the southern half of the jet precesses 
around an axis misaligned by 10$^\circ$ relative to the symmetry axis of
the northern jet. Such a model represents a somewhat better fit to the positions 
of the southern HH groups, and ascribes the same age to the B and E groups 
(4,900 yr) and to the C and F groups (6,900 yr). Moreover, the bisector of segments 
joining pairs of same-age positions on either sides of the jet are progressively 
displaced towards the east as larger ages are considered. In particular, the 
bisectors at ages 4,900 yr and 6,900 yr are very nearly coincident with the 
measured bisectors of the B-E and C-F pairs (Fig.\ 3b).

The model showed in Fig.\ 3b certainly represents an oversimplification 
of the true structure of the jet in PV Cep, and is clearly not unique. It does
demonstrate, however, that a slightly asymmetric jet can provide a better 
description of the structure of the flow than a symmetric one, and could 
readily explain the bisector results of GA2004 with no need for a large
stellar velocity. It would be interesting to investigate more realistic
numerical models of bipolar jets encountering asymmetric surroundings.
In particular, it would be useful to know in what domain of the parameter 
space they can produce bisector results similar to those found in PV Cep
by GA2004. 

It is worth mentioning that the gas immediately south of PV Cep
appears to be significantly denser than the gas immediately north of it
(Arce \& Goodman 2002a, 2002b). Thus, one might expect the southern 
lobe to be slower than the northern lobe and not the other way around, 
as our asymmetric model would suggest. The distribution of the gas 
surrounding PV Cep, however, is likely to be quite asymmetric. For 
instance, knot C (to the north) is known to entrain a massive molecular 
gas shell, whereas no entrained gas is associated with its southern 
counterpart knot F (Arce \& Goodman 2002a, 2002b). This suggests that, 
on large scales, the density of molecular material north of PV Cep might 
be higher that south of it. This is reversed compared to the density distribution 
in the immediate surroundings of PV Cep. 

In addition to their bisector results, GA2004 mention two pieces of evidence
supporting the idea that PV Cep might be a run-away star. The first one is
the existence of a tail of red-shifted gas located to the east of PV Cep. The
second is the wiggling of the southern jet seen at small scales. These two 
effects have not been reported in other outflow systems, and it is unclear if
they have a direct relation to the asymmetry of the jet in PV Cep. It certainly
would be worthwhile to search for these effects in other outflow systems 
(symmetric or not) and to investigate numerically in what kind of situation
they would tend to develop.

\section{Conclusions}

Comparing two VLA observations separated by 10.5 years, we have
measured the proper motion of the young star PV Cep. This measurement
shows that PV Cep is essentially stationary with respect to its
surroundings, and is not a fast-moving run-away star, as suggested by
GA2004. The jet morphological characteristics that led GA2004 to conclude 
that PV Cep might be a run-away star might instead result from an asymmetry 
in the jet itself. GA2004 had built a case against PV Cep reflected in the title of
their paper: ``PV Cephei: young star caught speeding?". Given the new 
evidence presented here, that accusation is now under question.

\acknowledgments LFR, LL, JC and ACR acknowledge the support of DGAPA,
UNAM, and of CONACyT (M\'exico).  L. G. was supported for this research through 
a stipend from the International Max Planck Research School (IMPRS) for Astronomy 
and Astrophysics at the Universities of Bonn and Cologne. This research has made 
use of the SIMBAD database, operated at CDS, Strasbourg, France.

\end{document}